\documentclass[%
 reprint,
 superscriptaddress,
 amsmath,amssymb,
 aps,
prb,
floatfix,
]{revtex4-2}

\usepackage{graphicx, epstopdf}
\usepackage{dcolumn}
\usepackage{bm}
\usepackage{gensymb}
\newcommand{\picWidth}{8.6cm}
\usepackage[colorlinks,linkcolor=blue,hyperindex,CJKbookmarks]{hyperref}
\makeatletter
\begin{document}
\title{
Spectra of a Gapped Quantum Spin Liquid with a Strong Chiral Excitation on the Triangular Lattice
}
\author{Ta Tang}
    \affiliation{Department of Applied Physics, Stanford University, California 94305, USA.}
\author{Brian Moritz}
    \affiliation{Stanford Institute for Materials and Energy Sciences, SLAC National Accelerator Laboratory, 2575 Sand Hill Road, Menlo Park, California 94025, USA.}
\author{Thomas P. Devereaux}
    \affiliation{Stanford Institute for Materials and Energy Sciences, SLAC National Accelerator Laboratory, 2575 Sand Hill Road, Menlo Park, California 94025, USA.}
    \affiliation{Department of Materials Science and Engineering, Stanford University, Stanford CA 94305.}
    \affiliation{Geballe Laboratory for Advanced Materials, Stanford University, Stanford, CA 94305, USA.}

\date{\today}

\begin{abstract}
    While a quantum spin liquid (QSL) phase has been identified in the $J_1$-$J_2$ Heisenberg model on a triangular lattice via numerical calculations, debate persists about whether or not such a QSL is gapped or gapless, with contradictory conclusions from different techniques. Moreover, information about excitations and dynamics is crucial for the experimental detection of such a phase. In this work, we use exact diagonalization to characterize signatures of a QSL phase on the triangular lattice through the dynamical spin structure factor $\mathcal{S}(q,\omega)$ and Raman susceptibility $\mathcal{\chi}(\omega)$. We find that spectra for the QSL phase show distinct features compared to those of neighboring phases; and both the Raman spectra and spin structure factor show gapped behaviour in the QSL phase. Interestingly, there is a prominent excitation mode in the Raman $A_2$ channel, indicating a strong subleading tendency toward a chiral spin liquid phase.
\end{abstract}

\pacs{Valid PACS appear here}
\maketitle

Quantum spin liquids (QSLs), characterized by the lack of magnetic order approaching zero temperature, were first considered by Anderson \cite{andersonResonatingValenceBonds1973} as an alternative ground state to the antiferromagnetic N\'eel phase. Later, QSLs were suggested as a possible route to high temperature superconductivity \cite{andersonResonatingValenceBond1987, baskaranResonatingValenceBond1987}, where preexisting singlet pairs may become superconducting upon doping. In addition to possible links with superconductivity, QSLs are massively entangled and can support exotic excitations, which can be utilized for topological quantum computation \cite{kitaevFaulttolerantQuantumComputation2003}. 

Among various lattices that have been suggested to host QSL phases, the triangular lattice plays an important role, as it was originally proposed by Anderson and many QSL candidates have this underlying lattice structure \cite{shimizuSpinLiquidState2003, kurosakiMottTransitionSpin2005}.
Although the Heisenberg model with a nearest neighbor interaction $J_1$ on the triangular lattice has been found to have long range antiferromagnetic order \cite{jolicoeurGroundstatePropertiesHeisenberg1990,bernuSignatureEelOrder1992, bernuExactSpectraSpin1994,capriottiLongRangeEelOrder1999}, adding longer-range interactions may increase frustration and help realize a QSL state. Numerical studies have reached a consensus that there is indeed a QSL phase on the triangular lattice with a next-nearest neighbor interaction $0.08\lesssim J_2/J_1 \lesssim 0.16$ \cite{zhuSpinLiquidPhase2015a,huCompetingSpinliquidStates2015,jiangNatureQuantumSpin2022, kanekoGaplessSpinLiquidPhase2014, iqbalSpinLiquidNature2016, wietekChiralSpinLiquid2017a}.
However, the nature of this QSL phase remains under active investigation, as some density matrix renormalization group (DMRG) calculations suggest that the QSL phase on the triangular lattice is a gapped spin liquid \cite{zhuSpinLiquidPhase2015a,huCompetingSpinliquidStates2015,jiangNatureQuantumSpin2022}, while variational quantum Monte Carlo (VMC) simulations \cite{kanekoGaplessSpinLiquidPhase2014,iqbalSpinLiquidNature2016} and a DMRG simulation with flux insertion \cite{huDiracSpinLiquid2019} suggest that the phase is a $U(1)$ gapless spin liquid .

While theoretical debates persist, tremendous progress has been made in the experimental identification of QSLs. 
Promising QSL candidates include triangular lattice systems such as 
$\kappa\text{-}\mathrm{(ET)_2X}$ \cite{shimizuSpinLiquidState2003,kurosakiMottTransitionSpin2005} and
$\mathrm{EtMe_3SB[Pd(dmit)_2]_2}$ \cite{itouQuantumSpinLiquid2008, yamashitaHighlyMobileGapless2010}. 
The lack of magnetic order down to the lowest accessible temperatures in these materials is a strong indication for the presence of a QSL ground state. However, critical questions remain about how to identify/distinguish experimentally between QSL phases, and how to link experimental measurements to theoretical models. 
While numerical methods like DMRG and VMC are powerful tools for studying ground state properties for large system simulations, it can be more difficult to study the dynamical properties of the system, and therefore difficult to provide results that can be compared directly to certain experimental measurements, such as the dynamical spin structure factor, as measured in neutron scattering, or the Raman spectra \cite{wulferdingRamanSpectroscopicDiagnostic2019a, broholmQuantumSpinLiquids2020, shaginyanTheoreticalExperimentalDevelopments2020}.

Here we study the $J_1$-$J_2$ Heisenberg model using exact diagonalization (ED) \cite{RevModPhys.66.763}, which exactly captures low lying eigenstates and can provide information about excitations and dynamics, albeit for small system sizes. 
Specifically, we obtain the dynamical spin structure factor $\mathcal{S}(q, \omega)$, which shows distinct features while tuning the ratio $J_{2}/J_{1}$, indicating the presence of distinct phases. We also extract the value of the spin excitation gap from finite-size scaling of $\mathcal{S}(q, \omega)$, which extrapolates to a finite value. 
In addition to $\mathcal{S}(q, \omega)$, the Raman spectrum also serves as an important experimental probe for QSLs \cite{wulferdingRamanSpectroscopicDiagnostic2019a}.
Here, we derive the lowest order Raman scattering operators for different symmetry channels and calculate the Raman susceptibility $\mathcal{\chi}(\omega)$ to characterize different phases. 
The QSL phase possesses distinct spectral features when compared to the nearby phases, tuning through the $J_{2}/J_{1}$ phase diagram.

The $J_1$-$J_2$ Heisenberg Hamiltonian is defined as
\begin{equation}
    H = \sum_{\left<ij\right>} J_1 \bm{S}_i \cdot \bm{S}_j + \sum_{\left<\left<ij\right>\right>} J_2 \bm{S}_i \cdot \bm{S}_j,
\end{equation}
where $\bm{S}_i = (S_i^x, S_i^y, S_i^z)$ denotes the spin vector on site i; $J_1$ is the nearest neighbor spin-exchange interaction and is set to 1; $J_2$ is the next nearest neighbor spin-exchange interaction; $\left<ij\right>$ denotes nearest-neighbor sites and $\left<\left<ij\right>\right>$ denotes next-nearest-neighbor sites.

\begin{figure}[tbph!]
    \begin{center}
        \includegraphics[width=\picWidth]{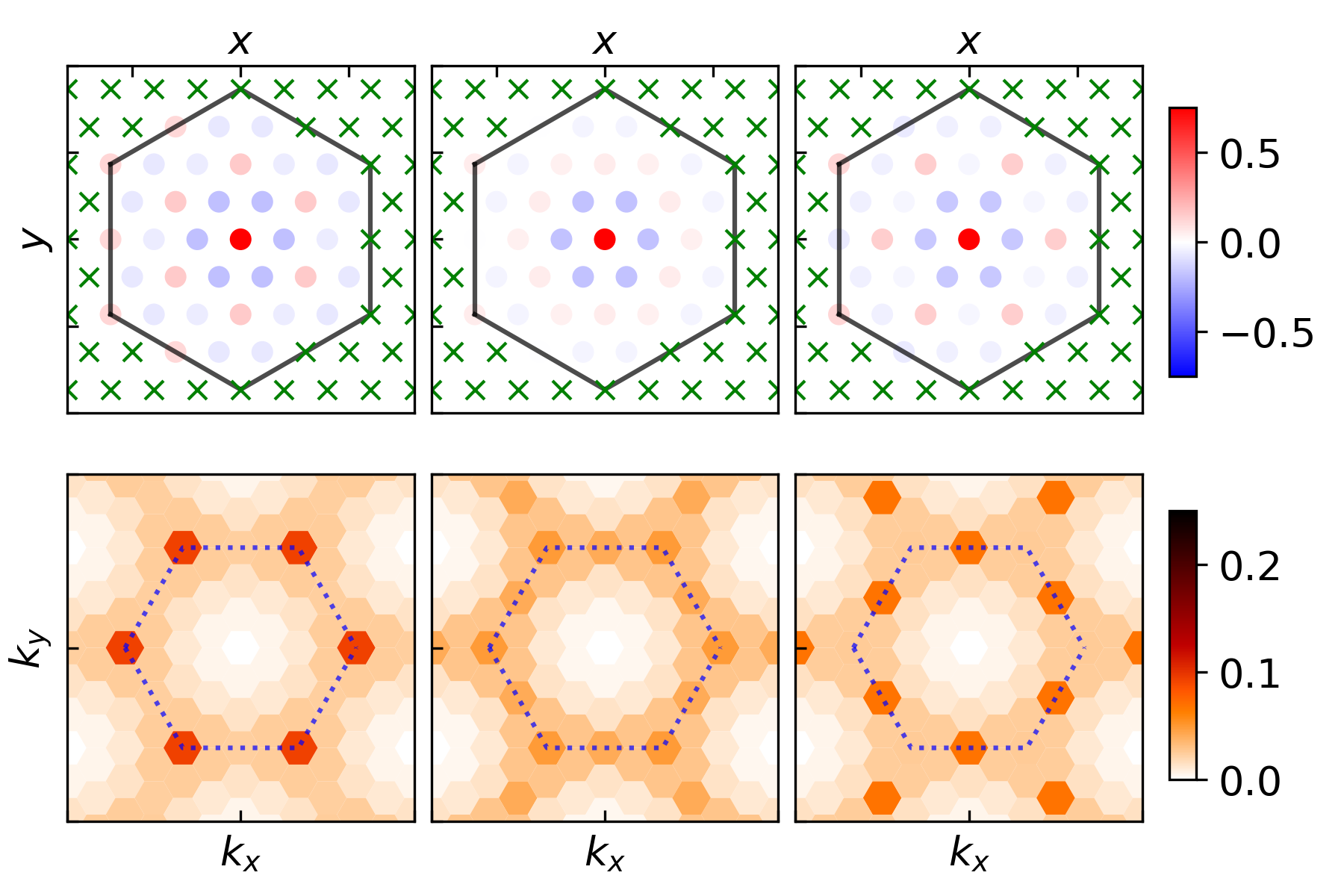}
    \end{center}
    \caption{
        Spin-spin correlation plotted in real space $\mathcal{S}_r$ (upper panel) and momentum space $\mathcal{S}_q$ (lower panel). The left column is for $J_2/J_1 = 0.0$, featuring peaks at BZ corners ($K$ points) for the $120\degree$ AF order. The right column with $J_2/J_1 = 0.24$ shows stripe order with peaks at the centers of the BZ edges ($M$ points); in the spin liquid phase parameter regime ($J_2/J_1 = 0.15$, middle column), spin-spin correlations decay very fast and the peaks at $K$ and $M$ points are largely suppressed and have similar intensity.
    }
    \label{pic:srsk}
\end{figure}

Previous numerical studies \cite{zhuSpinLiquidPhase2015a, huCompetingSpinliquidStates2015, kanekoGaplessSpinLiquidPhase2014, iqbalSpinLiquidNature2016, wietekChiralSpinLiquid2017a} have established that for small $J_2/J_1$, the system is in a $120\degree$ antiferromagnetically ordered state (hereafter $120\degree$ AF). Increasing $J_2 ~(\gtrsim 0.08J_1)$, the system transitions into a QSL phase characterized by exponentially vanishing spin-spin correlations. For larger $J_2 ~(\gtrsim 0.16J_1)$, the system is in a two-sublattice striped phase. In Fig.~\ref{pic:srsk}, we show the spin-spin correlations and the static spin structure factor obtained using ED on a 36-site cluster for three different values of $J_2$, representing the three phases. The spin-spin correlations are defined as  
\begin{equation}
    \mathcal{S}_r = \frac{1}{N}\sum_{i = 0}^{N - 1}\left<\bm{S}_{r_i}\cdot\bm{S}_{r_i + r}\right>,
\end{equation}
and the static spin structure factor is obtained by Fourier transforming $\mathcal{S}_r$
\begin{equation}
    \mathcal{S}_q = \frac{1}{N}\sum_{i = 0}^{N - 1}\mathcal{S}_{r_i}\mathrm{exp}(i\bm{q}\cdot\bm{r}_i),
\end{equation}
where N is the number of sites.
For $J_2/J_1 = 0$, there are prominent peaks at the Brillouin zone (BZ) corners ($K$ points), indicating the $120\degree$ AF order; large $J_2/J_1$ yields $\mathcal{S}_q$ peaks at the BZ edges ($M$ points), characteristic of stripe order. In the intermediate region $0.08\lesssim J_2/J_1 \lesssim 0.16$, a ring of peaks around the BZ boundary form, where the intensity at the $K$ and $M$ points is comparable, yet largely suppressed in comparison to the ordered states. The rapid decay of the real-space spin-spin correlations in the intermediate phase also serves as an indication of the QSL phase in this parameter regime.
These results are consistent with previous ED  \cite{bernuSignatureEelOrder1992, wietekChiralSpinLiquid2017a} and DMRG \cite{zhuSpinLiquidPhase2015a,huCompetingSpinliquidStates2015,jiangNatureQuantumSpin2022} studies. 

Next, signatures of a gap in the excited state spectra of the various phases are investigated via the dynamical spin structure factor and the polarization-dependent Raman susceptibilities.

\begin{figure}[tbph!]
    \begin{center}
        \includegraphics[width=\picWidth]{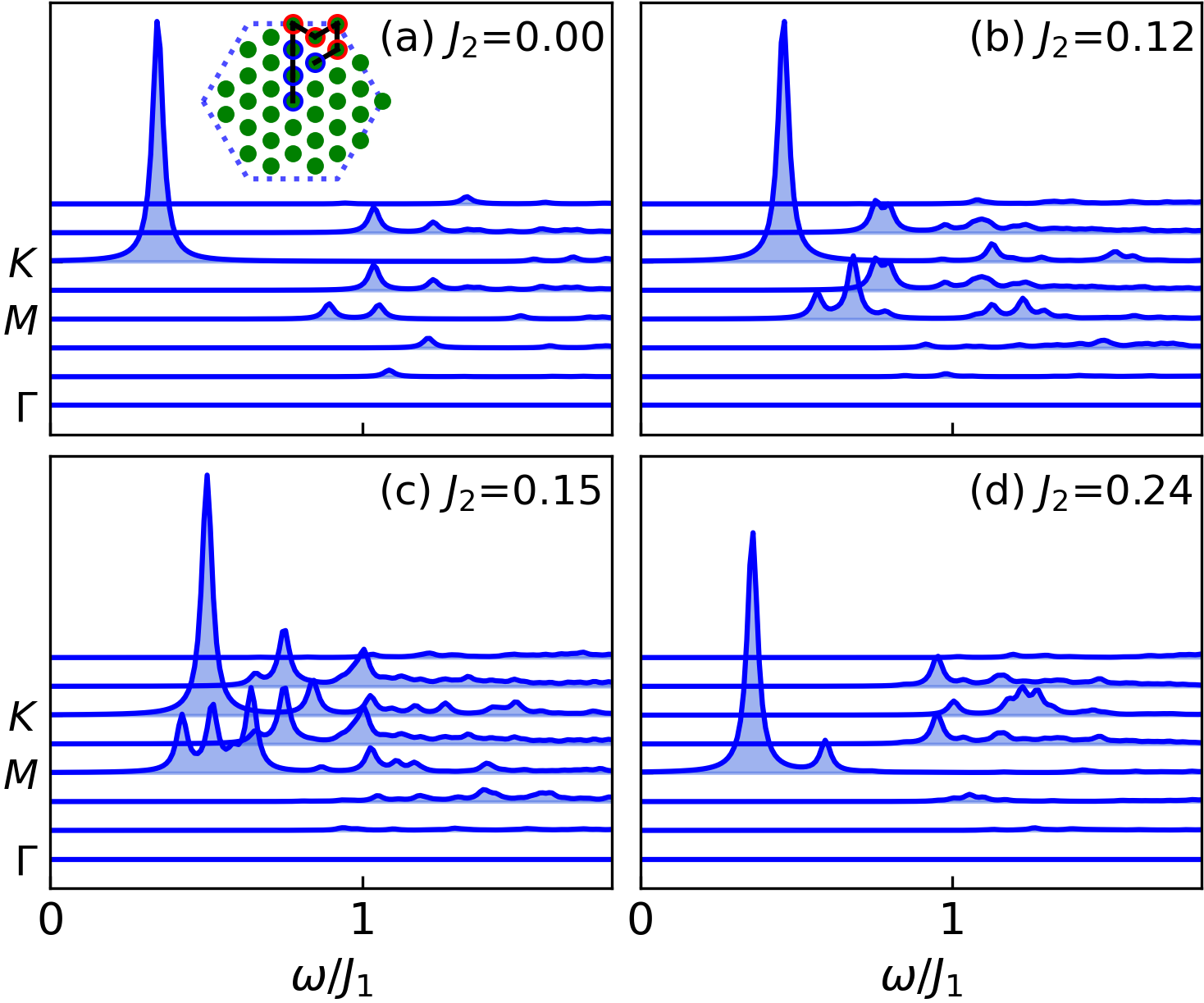}
    \end{center}
    \caption{
        Dynamical spin structure factor at representative momentum points in three different phases: (a) $J_2/J_1 = 0$ in the $120^o$ AF phase, (b)(c) $J_2/J_1 = 0.12, 0.15$ in the spin liquid phase, and (d) $J_2/J_1 = 0.24$ in the stripe phase. The inset shows the path in momentum space starting at $\Gamma$ point for drawing $\mathcal{S}^{zz}(q, \omega)$. $\Gamma$ is the BZ center, $M$ is the BZ edge center, and $K$ is the BZ corner. Momentum points near the Brillouin zone boundary are colored in red.
    }
    \label{pic:sqw}
\end{figure}

The dynamical spin structure factor is defined as
\begin{equation}
    \mathcal{S}^{zz}(q, \omega) = -\frac{1}{\pi}\mathrm{Im}\left<G\right|S^z_{-q}\frac{1}{\omega + E_0 - H + i\eta}S^z_q\left|G\right>,
\end{equation}
where $S^z_q\!=\!\frac{1}{\sqrt{N}} \sum_{i = 0}^{N - 1}S^z_{r_i}\mathrm{exp}(i\bm{q}\cdot\bm{r}_i)$, $\left|G\right>$ is the ground state and $E_0$ is the ground state energy. 

$S^{zz}(q, \omega)$ for different values of $J_2$ representing the three different phases are shown Fig.~\ref{pic:sqw}. $S^{zz}(q, \omega)$ for different momentum points in each plot from bottom to top correspond to those labeled by the path starting from the $\Gamma$ point in the inset of Fig.~\ref{pic:sqw}(a). All other momentum points in the BZ are related to these points through rotation or reflection symmetries.
In the $120\degree$ AF phase, the lowest excitation is at the ordering wave vector $K$. Increasing $J_2$, the excitation gap at $K$ increases while the excitation gap at $M$ and other momentum points near the BZ boundary shrink. In the spin liquid phase, the excitation gaps become comparable at $K, M$, and other points near the BZ boundary [colored in red in the inset of Fig.~\ref{pic:sqw}(a)]. The gap at $M$ becomes the smallest once the system enters the stripe phase when further increasing $J_2$. 

These three phases exhibit distinct spin excitation spectral features on finite clusters using ED. We note that our result qualitatively agrees well with that obtained by a dynamical variational Monte Carlo approach \cite{ferrariDynamicalStructureFactor2019}. 
In the thermodynamic limit, the two ordered phases are expected to become gapless, having gapless excitations emanating from the ordering wave vectors in $S^{zz}(q, \omega)$. 
However, whether the QSL phase is gapped or gapless remains unclear. 

Spin excitation gaps extracted from $S^{zz}(q, \omega)$ may be compared for different simulation cluster sizes to extrapolate to a thermodynamic limit. Unfortunately, only two clusters (12-site and 36-site) are available with $D_6$ point group symmetry and $S_z^{tot} = 0$ for reasonable computational cost. The next larger system size would be 48-sites, which is near the computational limit for ED. Therefore, it is hard to truly extrapolate to the 2-dimensional thermodynamic limit using ED.

In order to perform finite-size scaling, we instead use a set of clusters with size $4\times L_x$, where $L_x = 4, 5, 6, 7, 8$. This set of clusters all have the $M$ point but lack the $K$ point. Therefore, they are not appropriate to capture the $120\degree$ AF phase and the spin gap extracted on these clusters for the $120\degree$ AF phase will have severe finite-size effects. Nevertheless, we can focus our attention on the behavior of the gap extracted in either the QSL phase or the stripe phase, depending on the Heisenberg exchange parameters. We note that although the other spatial direction only has 4 sites, finite size effects may be not severe as the spin-spin correlations decay quickly in the QSL phase, as observed in DMRG results with more sites in the other spatial direction \cite{zhuSpinLiquidPhase2015a,huCompetingSpinliquidStates2015}.

\begin{figure}[htpb!]
    \begin{center}
        \includegraphics[width=0\picWidth]{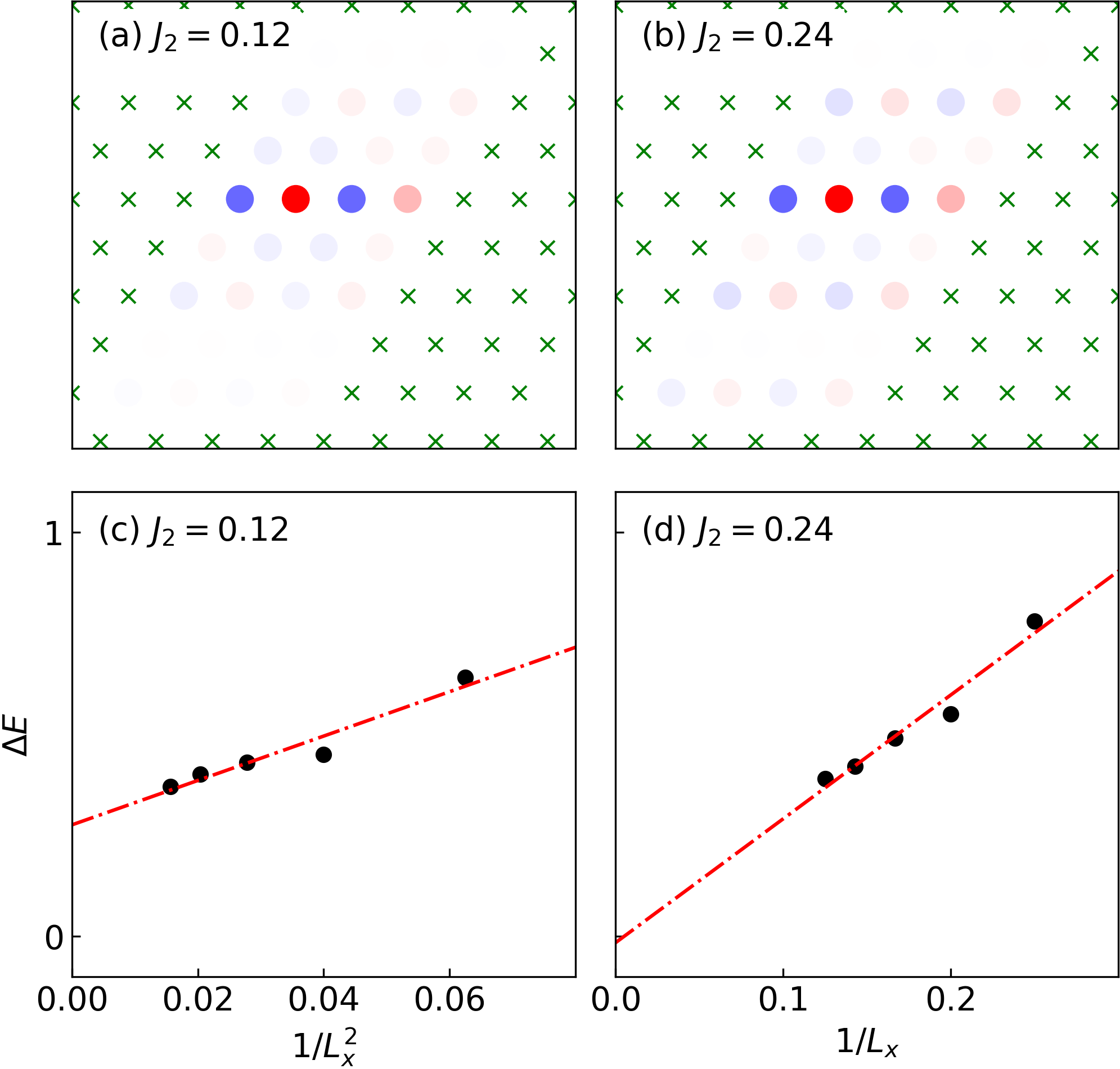} 
    \end{center}
    \caption{
        (a) and (b) are the spin-spin correlations for the $4 \times 8$ cluster in real space. (c) and (d) are the finite size scaling of the spin excitation gaps extracted from $S(q,\omega)$ on $4 \times L_x$ clusters. In the stripe phase (right column), the spin-spin correlation shows stripe order along the y direction, and the spin gap approaches 0 when $L_x\to\infty$ using a linear fit $\Delta E \sim 1/L_x$. In the spin liquid phase parameter regime (left column), the spin-spin correlation decays very fast along the y direction, and the system remains gapped with $\Delta E \approx 0.28J_1$ when $L_x\to\infty$ using a quadratic fit $\Delta E \sim 1/L_x^2$. It is also gapped with $\Delta E \approx 0.1J_1$ if using a linear fit. 
    }
    \label{pic:gap}
\end{figure}

In Fig.~\ref{pic:gap} (lower panel), the spin gap denoted as $\Delta E$ is extracted for $4\times L_x$ clusters with $J_2/J_1 = 0.12$, deep in the spin liquid phase, and $J_2/J_1 = 0.24$, well into the striped phase. Fitting $\Delta E \approx c\cdot k_x + \delta$ (a gapless linear spin wave), we see that the gap remains finite $\Delta E \approx 0.1J_1$ in the QSL phase, while the gap tends to zero for the striped phase. Fitting with quadratic dispersion ($\Delta E \approx c\cdot k_x^2 + \delta$) gives an even larger gap $\Delta E \approx 0.28J_1$ in the QSL phase as $k_x \rightarrow 0$. The extrapolated spin gap using the quadratic fit is consistent with the result from DMRG \cite{zhuSpinLiquidPhase2015a,huCompetingSpinliquidStates2015} on cylindrical clusters, which favour a gapped QSL.

The dynamical spin structure factor shows distinct features for the QSL phase and its neighboring phases. However, neutron scattering experiments usually require large samples or accumulating measurements from many samples to obtain a sizable signal \cite{wulferdingRamanSpectroscopicDiagnostic2019a}. This may hinder use of the dynamical spin structure factor to diagnosis a QSL experimentally. In contrast, inelastic light scattering can often yield larger scattering intensity from smaller samples and can be much easier to obtain experimentally. In the next section, we explore Raman scattering for the QSL and its neighboring phases.

In addition to neutron scattering, Raman spectroscopy serves as an important experimental probe~\cite{devereauxInelasticLightScattering2007,koRamanSignatureDirac2010,vernayRamanScatteringTriangular2007a,
perkinsRamanScatteringHeisenberg2008a,
knolleRamanScatteringSignatures2014,hassanRamanScatteringProbe2018,wulferdingRamanSpectroscopicDiagnostic2019a}. While $\mathcal{S}(q,\omega)$ probes excitations induced by flipping one spin ($\Delta S_z=\pm 1$), Raman scattering processes involve even numbers of spin flips ($\Delta S_z=0,\,\pm 2$). Furthermore, the scattering geometry (in-coming and out-going polarization discrimination) may be used to probe different symmetry channels. Thus Raman spectroscopy is capable of providing extra information about excitations and the interplay between lattice symmetry and underlying order. First, we derive the lowest order Raman scattering operators in different symmetry channels. Using these, we compute and compare the Raman susceptibility on the 36-site cluster for the distinct phases outlined in the previous analysis.

We can obtain the effective scattering operator in the spin basis by first considering Raman scattering (a photon-in/photon-out process) using the underlying light-matter interaction Hubbard Hamiltonian, and then project out double occupancies~\cite{devereauxInelasticLightScattering2007,koRamanSignatureDirac2010}. 

Since the Raman scattering operator contains dot products with the incoming and outgoing light polarizations(see supplementary material), it can be written in the general form
\begin{equation}
    \hat{M} = \sum_{\alpha\beta}M_{\alpha\beta}e_f^\alpha e_i^\beta,
\end{equation}
where $\bm{e}_i$ and $\bm{e}_j$ are the incoming and outgoing light polarization, and $\alpha$, $\beta$ denote spatial basis. 

We focus on the 36-site cluster with $D_6$ point group symmetry. The Raman scattering operator can be decomposed according to the irreducible representations of the $D_6$ symmetry group
\begin{eqnarray}
    \label{eq:ramanD6}
    \hat{M} & = & \hat{O}_{A_1}(e_f^x e_i^x+e_f^y e_i^y) + \hat{O}_{A_2}(e_f^x e_i^y - e_f^y e_i^x) \nonumber\\ 
            & + & \hat{O}_{E_2^{(1)}}(e_f^x e_i^x - e_f^y e_i^y) + \hat{O}_{E_2^{(2)}}(e_f^x e_i^y + e_f^y e_i^x),
\end{eqnarray}
where $A_1$, $A_2$ and $E_{2}^{(1/2)}$ denote different symmetry channels. 

To lowest order $O(t_1^2/U + t_2^2/U)$, where $t_1$ and $t_2$ denotes nearest neighbor and next nearest neighbor hopping integrals respectively, we obtain the Elliot-Fleury-Loudon scattering operator \cite{elliottPossibleObservationElectronic1963, fleuryScatteringLightOne1968}
which consists of terms $S_i \cdot S_j$
\begin{equation}
    \hat{M}_{EFL} \propto \sum_{\bm{r},\bm{r'}}\frac{2t_{\bm{r}\bm{r'}}^2}{U-\omega_i}(\bm{e}_i\cdot\bm{\delta})(\bm{e}_f\cdot\bm{\delta})\bm{S_r}\cdot\bm{S_r'},
\end{equation}
where $t_{\bm{r}\bm{r'}}$ is the hopping between site $\bm{r}$ and site $\bm{r'}$, $\bm{\delta} = \bm{r'} - \bm{r}$, $U$ is the onsite repulsion and $\omega_i$ is the incident photon energy.
We note that the derivation of $\hat{M}_{EFL}$ involves the same two-step virtual hoppings in the derivation of the Heisenberg Hamiltonian from the Hubbard Hamiltonian, and we have $J_1\sim 4t_1^2/U$ and $J_2\sim 4t_2^2/U$.
Elliot-Fleury-Loudon scattering operator gives the lowest order expressions for $\hat{O}_{A_1}$, $\hat{O}_{E_2^{(1)}}$ and $\hat{O}_{E_2^{(2)}}$. However, we note that the $A_2$ channel changes sign under reflection, but the Elliot-Fleury-Loudon term is invariant under reflection. Thus, $\hat{O}_{A_2}$ vanishes at this order. The lowest order non-zero $\hat{O}_{A_2}$ would be found at $O(t_1^3t_2/U^3)$, and consists of a sum of chiral terms $\bm{S_{r_1;r_2;r_3}} = \bm{S_{r_1}}\cdot (\bm{S_{r_2}}\times \bm{S_{r_3}})$. The derivation and specific forms for these scattering operators are given in the supplementary material.

It is helpful for the understanding of the Raman susceptibility to first look at the ground state point group symmetry across the three phases when tuning the ratio $J_2/J_1$.
From the eigenvalue spectrum (see the supplementary material or Ref.~\cite{wietekChiralSpinLiquid2017a}), we know that the transition from the QSL phase to the striped phase is related to a level crossing in the ground state. In the QSL phase, the ground state belongs to the symmetry subgroup $\Gamma.A_1$, where $\Gamma$ denotes momentum $0$ and $A_1$ denotes the trivial representation of the $D_6$ point group. The first excited state belongs to $\Gamma.E_2$ and is doubly degenerate, since $E_2$ is the two dimensional representation of the $D_6$ point group. Increasing $J_2$, the energy of the $\Gamma.E_2$ doublet goes down and eventually crosses $\Gamma.A_1$ at the phase transition between the QSL phase and the striped phase. 
Beyond the level crossing, the $\Gamma.A_1$ state energy remains slightly above that of $\Gamma.E_2$ in the striped phase, and as we will see, contributes to the very low frequency peak in the two $E_2$-channel Raman susceptibilities.
In contrast, the ground state in both the $120\degree$ AF phase and the QSL phase belongs to the $\Gamma.A_1$ sector and there is no level crossing.


\begin{figure}[t!]
    \begin{center}
        \includegraphics[width=8.5cm]{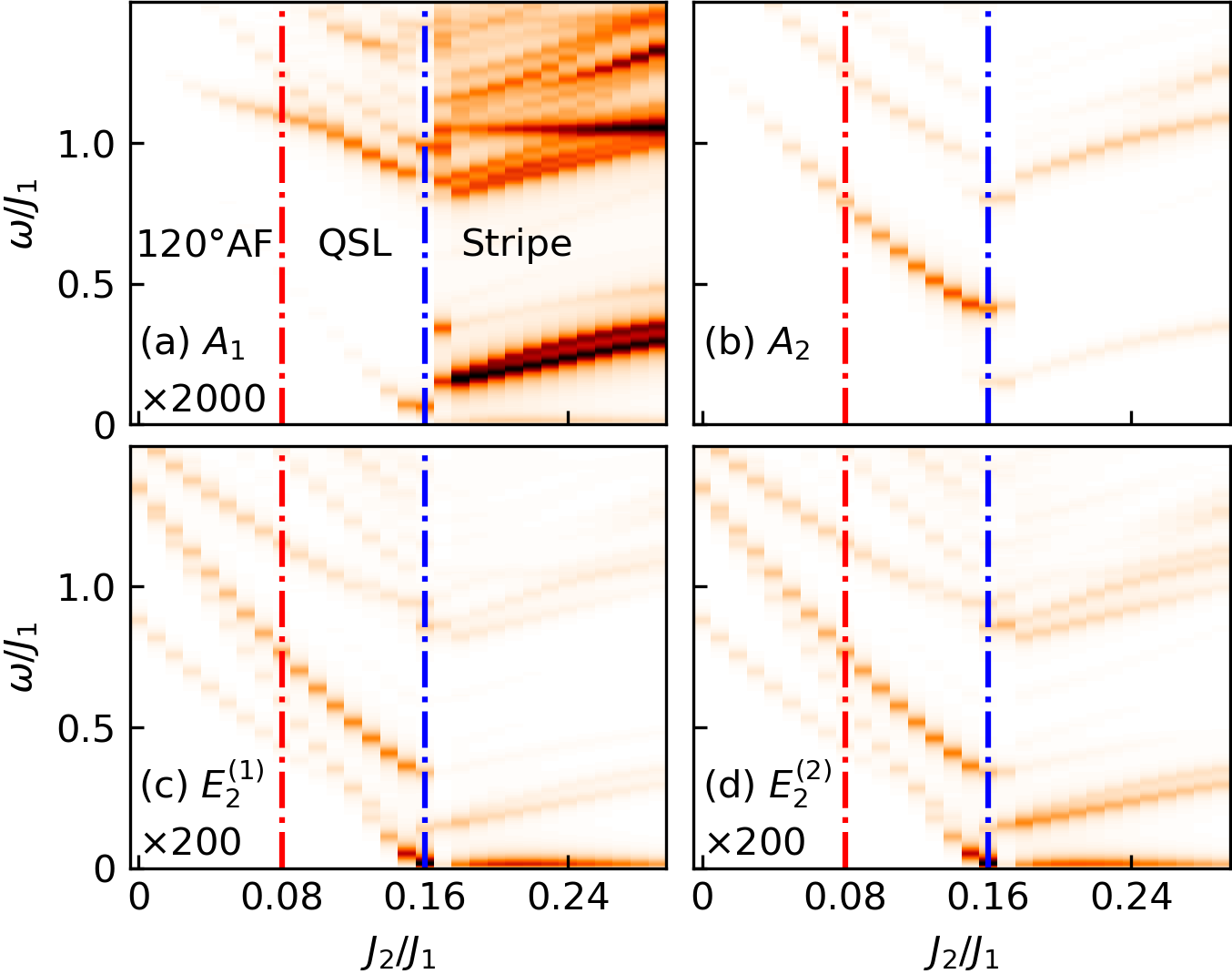}
    \end{center}
    \caption{
        Raman susceptibility for different symmetry channels as a function of $J_2/J_1$. 
        We do not include the $(t_1/U)^4$ factor in the $A_2$ channel spectra. All spectra are normalized to the maximum value of $A_2$ spectra. We enhanced the $A_1$ and $E_2$ spectra to make them visible.  
        The red and blue dashed line, corresponding to $J_2=0.08$ and $J_2=0.16$, mark the boundary between the $120\degree$ AF phase, the QSL phase and the stripe phase.
        In the stripe phase, the low excitation peaks in the $E_2$ channel originates from an excited state nearly degenerate to the ground state.
    }
    \label{pic:raman}
\end{figure}

The Raman spectra are obtained using
\begin{equation}
    R^\alpha(\omega)=-\frac{1}{\pi}\text{Im}\left<G\left|\hat{O}_\alpha^\dagger\frac{1}{\omega+E_0+i\epsilon-\hat{H}}\hat{O}_\alpha\right|G\right>,
\end{equation}
where $\hat{O}_\alpha$ denotes a Raman scattering operator in channel $\alpha$.
The Raman scattering susceptibility is defined as
\begin{equation}
    \chi^\alpha(\omega) = R^\alpha(\omega) - R^\alpha(-\omega),
\end{equation}
which removes the elastic peak in $R^\alpha(\omega)$.

In Fig.~\ref{pic:raman}, we plot $\chi^\alpha(\omega)$ as a function of $J_2/J_1$. 
$\chi^\alpha(\omega)$ changes dramatically when transitioning from the QSL phase to the striped phase as expected. 
As mentioned before, there is a very low energy peak in the striped phase $E_2$ channel susceptibility coming from the $\Gamma.A_1$ state.
Because only the $E_2$ channel Raman scattering operators connect the $\Gamma.E_2$ ground state with the $\Gamma.A_1$ excited state, this low energy peak is visible only in the $E_2$ Raman scattering channel.
In addition to this low frequency peak, compared to the QSL phase there are two strong peaks close to each other below $0.5\,J_1$ in the striped phase $A_1$ channel; they also are visible in $E_2$, but weaker. In the $A_2$ channel, there is a strong peak around $0.5\,J_1$ in the QSL phase, while there is a much weaker peak below $0.5\,J_1$ in the striped phase.

In contrast to the discontinuity observed in $\chi^\alpha(\omega)$ caused by the first order phase transition from the QSL phase to the striped phase, $\chi^\alpha(\omega)$ changes smoothly from the $120\degree$ AF phase into the QSL phase. However, we see that there is a level crossing for the lowest energy excitation in the $E_2$ channels, which occurs with the phase transition from the $120\degree$ AF phase into the QSL phase. This signals a second order phase transition. Simultaneously, we see that a low frequency peak below $0.5\,J_1$ develops in the $A_1$ channel after entering the QSL phase. These distinct features in the Raman spectrum for different phases may be utilized to identify them experimentally.

We note that since the $A_2$ channel scattering operator is derived from higher order terms, it has a $(t_1/U)^2$ prefactor compared to scattering operators from other channels. Consequently, the spectral intensity will scale relatively as $(t_1/U)^4$ compared to other channels. In Fig.~\ref{pic:raman}, we do not include this prefactor in the $A_2$ channel spectra, so one can compare the relative intensity of excitations induced by scattering operators coming from different symmetry as if they are treated on the same order. This is helpful to truly identify the dominant excitations. As we can see, the $A_2$ channel excitation is dominant and is especially strong in the QSL phase. The intensity of features in the $A_1$ and $E_2$ channels of Fig.~\ref{pic:raman} is multiplied by $2000$ and $200$, respectively, so as to make them visible in comparison to the intensity of features in the $A_2$ channel. This reveals that the dominant excitation in the QSL phase is a chiral mode and the system may have a strong subleading tendency toward chiral order.
Tendency towards chirality in the $J_1$-$J_2$ model was analyzed in an early work by Baskaran \cite{baskaranNovelLocalSymmetries1989}. However, recent works have suggested that a $J_2$ term alone is not enough to break time reversal symmetry in the ground state.
Wietek et al.~\cite{wietekChiralSpinLiquid2017a} suggested that in the QSL phase parameter regime, a chiral spin liquid (CSL) phase can be realized by adding a small chiral term via a magnetic field. 
Alternatively, an additional four-spin ring exchange interaction originating from the underlying Hubbard Hamiltonian is also suggested to realize a CSL \cite{cookmeyerFourSpinTermsOrigin2021, szaszChiralSpinLiquid2020}.

In summary, we obtained the dynamical spin structure factor and Raman spectra for the QSL phase, and its neighboring phases, on the triangular lattice $J_1$-$J_2$ Heisenberg model. 
For the $120\degree$ AF phase and the striped phase, the lowest excitation in $\mathcal{S}(q,\omega)$ occurs at the corresponding wave vector and should become gapless in the thermodynamic limit. In contrast, the spin gap in the QSL phase extrapolates to a finite value using the $4\times L_x$ clusters, in agreement with DMRG results~\cite{zhuSpinLiquidPhase2015a,huCompetingSpinliquidStates2015,jiangNatureQuantumSpin2022}. We also find that the lowest spin excitations in the QSL phase spread accross the entire Brillouin zone boundary, with gap sizes that are comparable on the 36-site cluster.
In the Raman spectra, we see a level crossing in the $E_2$ channel and a low energy peak below $0.5\,J_1$ that gradually develops in the $A_1$ channel as the system transitions from the $120\degree$ AF phase to the QSL phase. There are abrupt changes in the spectra that occur in all channels transitioning from the QSL phase to the striped phase.

We note that there is no sign of a gappless continuum~\cite{wulferdingRamanSpectroscopicDiagnostic2019a} in any Raman channel in the QSL phase. 
Combining results from the dynamical spin structure factor and Raman susceptibility (Elliot-Fleury-Loudon terms), the QSL phase is gapped in the $\Delta S_z = 0, \, \pm 1, \, \pm 2$ spin excitation channels, stronly suggestive of a gapped QSL phase. Interestingly, the Raman $A_2$ channel scattering operator consists of chiral terms and its spectra show a very prominent, but gapped mode across the QSL phase, suggesting a strong subleading tendency toward chiral order. Because the excitation in the $A_2$ channel is so strong in the QSL phase, it still can have comparable strength relative to other Raman channels, even accounting for all prefactors, making it easier to access experimentally. 

The distinct spectral features for different phases can serve as a fingerprint for identifying QSL signals in experiments on triangular lattice materials. 
One also can apply isotropic pressure or strain; and since $J_2$ falls-off faster as we increase lattice spacing, the $J_2/J_1$ ratio can be tuned in this fashion. The ratio also will be different in different materials; and one may tune $J_2/J_1$ through synthesis to observe the evolution of spectral features.

The authors would like to thank Yifan Jiang, Hongchen Jiang and Johannes Motruk for helpful discussions and suggestions. This work was supported by the U.S. Department of Energy, Office of Basic Energy Sciences, Division of Materials Sciences and Engineering, under Contract No.~DE-AC02-76SF00515. The computational results utilized the resources of the National Energy Research Scientific Computing Center (NERSC) supported by the U.S. Department of Energy, Office of Science, under Contract No.~DE-AC02-05CH11231.

\section*{supplementary Materials}
\subsection*{Eigenvalue Spectrum}
\begin{figure}[htpb!]
    \begin{center}
        \includegraphics[width=8.5cm]{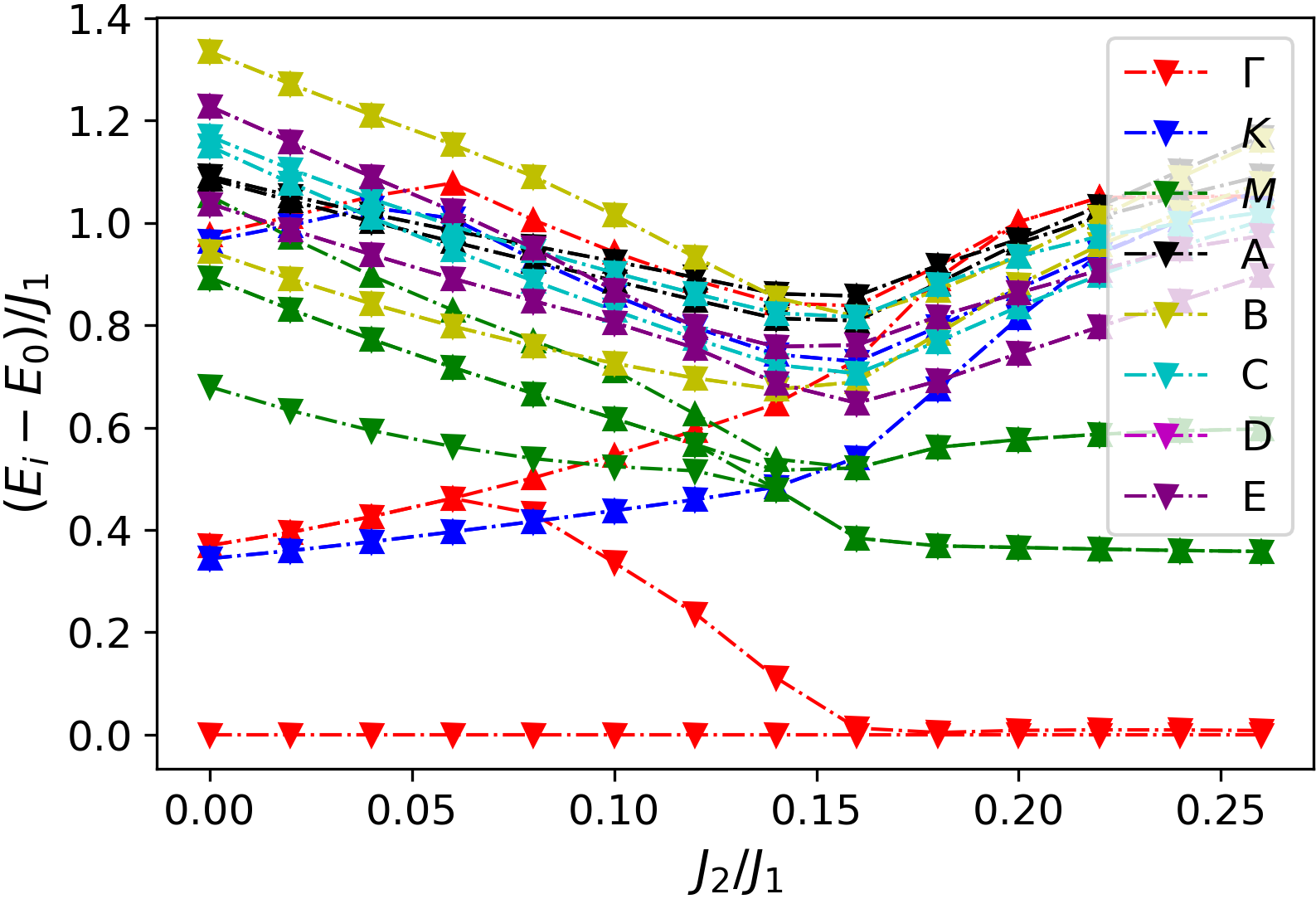}
    \end{center}
    \caption{
        Eigenvalues of the 36-site cluster as functions of $J_2/J_1$. Different colors represent different momentum subspaces.
        Here $\Gamma$ denotes the BZ center, $K$ dentotes the BZ corner, $M$ denotes the BZ edge center, and $A-E$ denotes moment points with increasing distance from the $\Gamma$ point. Downward triangles represent $S_z = 0$ states and upward triangles represent $S_z = 1$ states. 
    }
    \label{pic:eigs}
\end{figure}

Due to translation symmetry, the Hilbert space can be decomposed into subspaces labeled by different momenta and the Hamiltonian is block diagonal in these subspaces. In Fig.~\ref{pic:eigs}, we plot eigenvalues for the $S_z=0$ and $S_z=1$ sectors in different momentum subspaces as a function of $J_2/J_1$. The ground state is always in the $\Gamma$ (total momentum 0) subspace. We can further decompose the total momentum 0 subspace using the $D_6$ point group symmetry. 
There is a level crossing around $J_2/J_1 \approx 0.17$, before which the ground state is in the $\Gamma.A_1$ sector, and after which the ground state is doubly degenerate and comes from the $\Gamma.E_2$ sector.   
Here, $A_1$ is the trivial representation of the $D_6$ point group; $E_2$ is the two dimensional representation of the $D_6$ point group and has two copies $E_2^{(1)}$ and $E_2^{(2)}$.

\subsection*{Raman Scattering Operators}

The light-mater interaction Hamiltonian is
\begin{equation}
    H = H_{el} + H_{\gamma} + H_{int},
\end{equation}
where $H_{el}$ is the electronic part modeled by a Hubbard Hamiltonian with on-site Coulomb repulsion, nearest neighbor and next-nearest neighbor hopping, given by
\begin{eqnarray}
    H_{el} & = & H_t + H_U \nonumber\\
          & = & \sum_{\left<ij\right>,\sigma}\!t_{1}c_{i\sigma}^{\dagger}c_{j\sigma} + \!\!\!\sum_{\left<\left<ij\right>\right>,\sigma}\!\!t_{2}c_{i\sigma}^{\dagger}c_{j\sigma} + U\sum_{i}n_{i\uparrow}n_{i\downarrow},
\end{eqnarray}
and $H_{\gamma} = \omega_{\gamma} n_{\gamma}$ is the free photon Hamiltonian. The light-matter interaction is obtained by using Peierls substitution $c_{i\sigma}^{\dagger}c_{j\sigma} \rightarrow c_{i\sigma}^{\dagger}c_{j\sigma}\text{exp}(\frac{ie}{\hbar c}\int_j^i\bm{A}\cdot \text{d}\bm{r})$ and expanding to second order in the vector potential
\begin{eqnarray}
    H_{int} & = & H_{int}^{(1)} + H_{int}^{(2)} \nonumber\\
     & = & \sum_{ij\sigma} t_{ij} c_{i\sigma}^\dagger c_{j\sigma} \bigg\{\frac{ie}{\hbar c} \bm{A}(\frac{\bm{x}_i+\bm{x}_j}{2})\cdot (\bm{x}_i-\bm{x}_j) \nonumber\\
     & & - \frac{e^2}{2\hbar^2 c^2}\left[\bm{A}(\frac{\bm{x}_i+\bm{x}_j}{2})\cdot (\bm{x}_i-\bm{x}_j)\right]^2\bigg\},
\end{eqnarray}
where $H_{int}^{(1/2)}$ represent terms which are first/second order in $\bm{A}$.
We can obtain the resonant scattering operator via a perturbative expansion
\begin{equation}
    \hat{M}_{R} = H_{int}^{(1)}W\sum_{n=0}^{\infty}\left(H_t W \right)^n H_{int}^{(1)},
\end{equation}
where $W = 1/(\xi_i-(H_{U}+H_\gamma)+i\eta)$, and $\xi_i$ is the initial state energy. We only consider intermediate states with one hole and one double occupancy, thus $W \approx 1/(\omega_i - U)$ in our calculation.

We will use the following spin operator identities valid for singly occupied states
\begin{equation}
    c^\dagger_\sigma c_{\sigma'} = \tilde{\chi}_{\sigma'\sigma} = \frac{1}{2} \delta_{\sigma'\sigma} + \bm{S} \cdot \bm{\tau}_{\sigma'\sigma},
\end{equation}
\begin{equation}
    c_\sigma c^\dagger_{\sigma'} = \chi_{\sigma\sigma'} = \frac{1}{2} \delta_{\sigma\sigma'} - \bm{S} \cdot \bm{\tau}_{\sigma\sigma'},
\end{equation}
\begin{equation}
    (\bm{a}\cdot\bm{\tau})(\bm{b}\cdot\bm{\tau}) = (\bm{a}\cdot\bm{b}) I + i(\bm{a}\times\bm{b})\cdot\bm{\tau}
\end{equation}
where $\bm{\tau}$ is pauli matrices and $\bm{S} = \frac{1}{2} c^\dagger_\sigma \bm{\tau}_{\sigma\sigma'} c_{\sigma'}$ is the spin operator.

For convenience, we define the following vectors using triangular lattice basis vectors $\bm{a}_1 = (1, 0)$ and $\bm{a}_2 = (1/2, \sqrt{3}/2)$ 
\begin{equation}
    \bm{a}_3 = \bm{a}_2 - \bm{a}_1, \bm{R}_1 = \bm{a}_1 - \bm{a}_3, \bm{R}_2 = \bm{a}_1 + \bm{a}_2, \bm{R}_3 = \bm{a}_2 + \bm{a}_3
\end{equation}
\subsubsection{Zeroth Order}
For the lowest order, we have two pathways for each bond
\begin{eqnarray*}
    T_{0, a} & = & (\bm{e}_f \cdot \bm{\delta}_{1,2})(\bm{e}_i \cdot \bm{\delta}_{2,1})\frac{it_{2,1} \cdot it_{1,2}}{(\omega_i - U)}(c^\dagger_1 c_2)(c^\dagger_2 c_1) \\
     & = & C_0 \mathrm{tr}(\chi_{2} \tilde{\chi}_{1}) \\
     & = & C_0 \mathrm{tr}(\frac{1}{4}I - (\bm{S}_2 - \bm{S}_1) \cdot \bm{\tau} - (\bm{S}_1 \cdot \bm{\tau}) (\bm{S}_2 \cdot \bm{\tau})) \\
     & = & 2C_0(\frac{1}{4} - \bm{S}_1 \cdot \bm{S}_2),
\end{eqnarray*}
where $t_{i,j}$ denotes the hopping between vertex $v_i$ and $v_j$, $\bm{\delta}_{i,j}$ denotes the vector pointing from $v_i$ to $v_j$, $C_0$ equals to $(\bm{e}_f \cdot \bm{\delta}_{1,2})(\bm{e}_i \cdot \bm{\delta}_{1,2}){t^2_{1,2}} / {(\omega_i - U)}$, and $(c^\dagger_ic_j)$ denotes summation over spin $\sum_\sigma c^\dagger_{i\sigma}c_{j\sigma}$.
The inverse path (just exchange the index 1 and 2) gives the same answer. Thus we reproduces the Elliot-Fleury-Loudon scattering operator 
\begin{equation}
    \hat{M}_{EFL} \propto \sum_{\bm{r},\bm{r}'}\frac{4t_{\bm{r}\bm{r}'}^2}{\omega_i - U}(\bm{e}_i\cdot\bm{\delta})(\bm{e}_f\cdot\bm{\delta})\bm(\frac{1}{4} - {\bm{S}_r}\cdot\bm{S}_{r'}),
\end{equation}
where $\bm{\delta} = \bm{r}' - \bm{r}$.
We can decompose it into $A_1$ and $E_2$ channels (ignoring the constant terms in the $A_1$ channel)
\begin{equation}
    \hat{O}_{A_1} = \frac{t_1^2}{U-\omega_i}\sum_{nn}\bm{S}_i\cdot\bm{S}_j + \frac{t_2^2}{U-\omega_i}\sum_{nnn}3\bm{S}_i\cdot\bm{S}_j.
\end{equation}

\begin{eqnarray*}
\hat{O}_{E_2^{(1)}} & = & \frac{t_1^2}{U-\omega_i}\sum_{r}(\bm{S_r}\cdot\bm{S_{r+a_1}} - \frac{1}{2}\bm{S_r}\cdot\bm{S_{r+a_2}} \\
 & & - \frac{1}{2}\bm{S_r}\cdot\bm{S_{r+a_3}}) + \frac{t_2^2}{U-\omega_i}\sum_{r}(\frac{3}{2}\bm{S_r}\cdot\bm{S_{r+R_1}} \\
 & & + \frac{3}{2}\bm{S_r}\cdot\bm{S_{r+R_2}} - 3\bm{S_r}\cdot\bm{S_{r+R_3}}).
\end{eqnarray*}

\begin{eqnarray*}
\hat{O}_{E_2^{(2)}} & = & \frac{t_1^2}{U-\omega_i}\sum_{r}\frac{\sqrt{3}}{2}(\bm{S_r}\cdot\bm{S_{r+a_2}} - \bm{S_r}\cdot\bm{S_{r+a_3}}) \\
& + & \frac{t_2^2}{U-\omega_i}\sum_{r}\frac{3\sqrt{3}}{2}(\bm{S_r}\cdot\bm{S_{r+R_2}} - \bm{S_r}\cdot\bm{S_{r+R_1}}).
\end{eqnarray*}

We now try to find the lowest order terms for the $A_2$ channel.

\subsubsection{First Order}

The first order terms are produced by paths forming a closed triangle, and it was shown that paths cancel each other in pairs in the $A_2$ channel~\cite{koRamanSignatureDirac2010}.

\subsubsection{Second Order}

If we only consider nearest neighbor hopping, the pathways involved in a parallelogram cancel out in the $A_2$ channel~\cite{koRamanSignatureDirac2010}. Thus we consider the next lowest order terms which involve three nearest neighbor hopping $t_1$ and one next-nearest neighbor hopping $t_2$. The relevant paths form a closed triangle by 4 vertexes $v_1 \to v_2 \to v_3 \to v_4 \to v_1$, involving three nearest neighbor edges and one next-nearest neighbor edge. For convenience, we label the edge connecting vertex $v_i$ and $v_{i+1}$ as $\bm{\delta}_i$ where $4+1$ is identified as 1. Without loss of generality, we can assume the first three edges are nearest neighbor edges and $\bm{\delta}_4$ is a next nearest neighbor edge. We first consider paths which originate from vertex $v_1$ along edge $\bm{\delta}_1$: $v_1 \to v_2$. There are in total 4 such pathways, and the first one is

\begin{eqnarray*}
    T_{2, a} & = & C_{2}(\bm{\delta}_1, \bm{\delta}_4) (c^\dagger_1 c_4) (c^\dagger_4 c_3) (c^\dagger_3 c_2) (c^\dagger_2 c_1) \\
    & = & C_{2}(\bm{\delta}_1, \bm{\delta}_4) \mathrm{tr}\{\chi_4 \chi_3 \chi_2 \tilde{\chi}_1\} \\
    & \to & i C_{2}(\bm{\delta}_1, \bm{\delta}_4) (S_{321} + S_{421} + S_{431} - S_{432}),
\end{eqnarray*}
where $C_{2}(\bm{\delta}_1, \bm{\delta}_2) = -(\bm{e}_i \cdot \bm{\delta}_1) (\bm{e}_f \cdot \bm{\delta}_2) t_1^3 t_2 / (\omega_i - U)^3$, $S_{ijk} = (\bm{S}_i \times \bm{S}_j) \cdot \bm{S}_k$, and "$\to$" in the last line means we only keep chiral terms. The other three pathways' contributions are
\begin{eqnarray*}
    T_{2, b} & = & C_{2}(\bm{\delta}_1, \bm{\delta}_3) (c^\dagger_4 c_3) (c^\dagger_1 c_4) (c^\dagger_3 c_2) (c^\dagger_2 c_1) \\
    & = & -C_{2}(\bm{\delta}_1, \bm{\delta}_3) \mathrm{tr}\{\tilde{\chi}_4 \chi_3 \chi_2 \tilde{\chi}_1\} \\
    & \to & -i C_{2}(\bm{\delta}_1, \bm{\delta}_3) (S_{321} - S_{421} - S_{431} + S_{432}),
\end{eqnarray*}

\begin{eqnarray*}
    T_{2, c} & = & C_{2}(\bm{\delta}_1, \bm{\delta}_3) (c^\dagger_4 c_3) (c^\dagger_3 c_2) (c^\dagger_1 c_4) (c^\dagger_2 c_1) \\
    & = & -C_{2}(\bm{\delta}_1, \bm{\delta}_3) \mathrm{tr}\{\tilde{\chi}_4 \chi_3 \chi_2 \tilde{\chi}_1\} \\
    & \to & -i C_{2}(\bm{\delta}_1, \bm{\delta}_3) (S_{321} - S_{421} - S_{431} + S_{432}),
\end{eqnarray*}

\begin{eqnarray*}
    T_{2, d} & = & C_{2}(\bm{\delta}_1, \bm{\delta}_2) (c^\dagger_3 c_2) (c^\dagger_4 c_3) (c^\dagger_1 c_4) (c^\dagger_2 c_1) \\
    & = & C_{2}(\bm{\delta}_1, \bm{\delta}_2) \mathrm{tr}\{\tilde{\chi_4} \tilde{\chi_3} \chi_2 \tilde{\chi}_1\} \\
    & \to & i C_{2}(\bm{\delta}_1, \bm{\delta}_2) (-S_{321} - S_{421} + S_{431} - S_{432}),
\end{eqnarray*}

Other pathways on this triangle can be generated by cyclic permutation or inverse of the vertex indexes: $(1 \to 2 \to 3 \to 4) \to (2 \to 3 \to 4 \to 1), (1 \to 2 \to 3 \to 4) \to (4 \to 3 \to 2 \to 1) \dots$ There are 8 such permutations and each one has 4 pathways as listed above. After summing them up, and considering all the possible triangles, we have the lowest order $A_2$ channel scattering operator
\begin{eqnarray*}
\hat{O}_{A_2} & = & \frac{t_1^3t_2}{(U-\omega_i)^3}\sum_{r} 4\sqrt{3} i (\bm{S_{r;r+R_2;r+a_1}}+\bm{S_{r;r+R_2;r+2a_1}} \\ 
& + & \bm{S_{r+a_1;r+R_2;r+2a_1}} + \text{Rotations} - \text{Reflections}) ,\\
\end{eqnarray*}
where $\bm{S_{r_1;r_2;r_3}} = \bm{S_{r_1}}\cdot (\bm{S_{r_2}}\times \bm{S_{r_3}})$, ``Rotations'' means terms obtained by rotating previous terms by $\pi/6,\dots, 5\pi/6$, and ``Reflections'' means terms obtained by reflecting previous terms along an axis. 

\bibliography{qsl}

\end{document}